\documentclass[aps,prappl,reprint,longbibliography,superscriptaddress]{revtex4-1}  
\usepackage{graphicx}  
\usepackage{dcolumn}   
\usepackage{multirow}
\usepackage{bm}        
\usepackage{amssymb}   
\usepackage{amsmath}   
\usepackage{siunitx}   
\usepackage{color}     

\usepackage[normalem]{ulem} 

\usepackage{blindtext}

\usepackage[colorinlistoftodos]{todonotes}
\usepackage{verbatim}
\usepackage[normalem]{ulem} 

\DeclareSIUnit\torr{Torr}
\DeclareSIUnit\sq{\ensuremath{\Box}}

\newcommand*\chem[1]{\ensuremath{\mathrm{#1}}}

\begin{document}


\title{Geometric Scaling of Two-Level-System Loss in Superconducting Resonators}

\author{David~Niepce}
\email{david.niepce@chalmers.se}
\affiliation{Chalmers University of Technology, Microtechnology and Nanoscience, SE-41296, Gothenburg, Sweden}
\author{Jonathan J.~Burnett}
\affiliation{National Physical Laboratory, Hampton Road, Teddington, Middlesex, TW11 0LW, United Kingdom}
\author{Mart\'{i}~Gutierrez~Latorre}
\affiliation{Chalmers University of Technology, Microtechnology and Nanoscience, SE-41296, Gothenburg, Sweden}
\author{Jonas~Bylander}
\affiliation{Chalmers University of Technology, Microtechnology and Nanoscience, SE-41296, Gothenburg, Sweden}
\vskip 0.25cm

\date{\today}

\begin{abstract}


We perform an experimental and numerical study of dielectric loss in superconducting microwave resonators at low temperature. 
Dielectric loss, due to two-level systems, is a limiting factor in several applications, e.g. superconducting qubits, Josephson parametric amplifiers, microwave kinetic-inductance detectors, and superconducting single-photon detectors.
Our devices are made of disordered NbN, which, due to magnetic-field penetration, necessitates 3D finite-element simulation of the Maxwell--London equations at microwave frequencies to accurately model the current density and electric field distribution. 
From the field distribution, we compute the geometric filling factors of the lossy regions in our resonator structures and fit the experimental data to determine the intrinsic loss tangents of its interfaces and dielectrics. 
We emphasise that the loss caused by a spin-on-glass resist such as hydrogen silsesquioxane (HSQ), used for ultrahigh lithographic resolution relevant to the fabrication of nanowires, and find that, when used, HSQ is the dominant source of loss, with a loss tangent of $\delta^i_{HSQ} = \SI{8e-3}{}$.

\end{abstract}

\maketitle

\section{Introduction}


\begin{figure*}
    \centering
    \begin{minipage}{\columnwidth}
        \includegraphics[width=7cm]{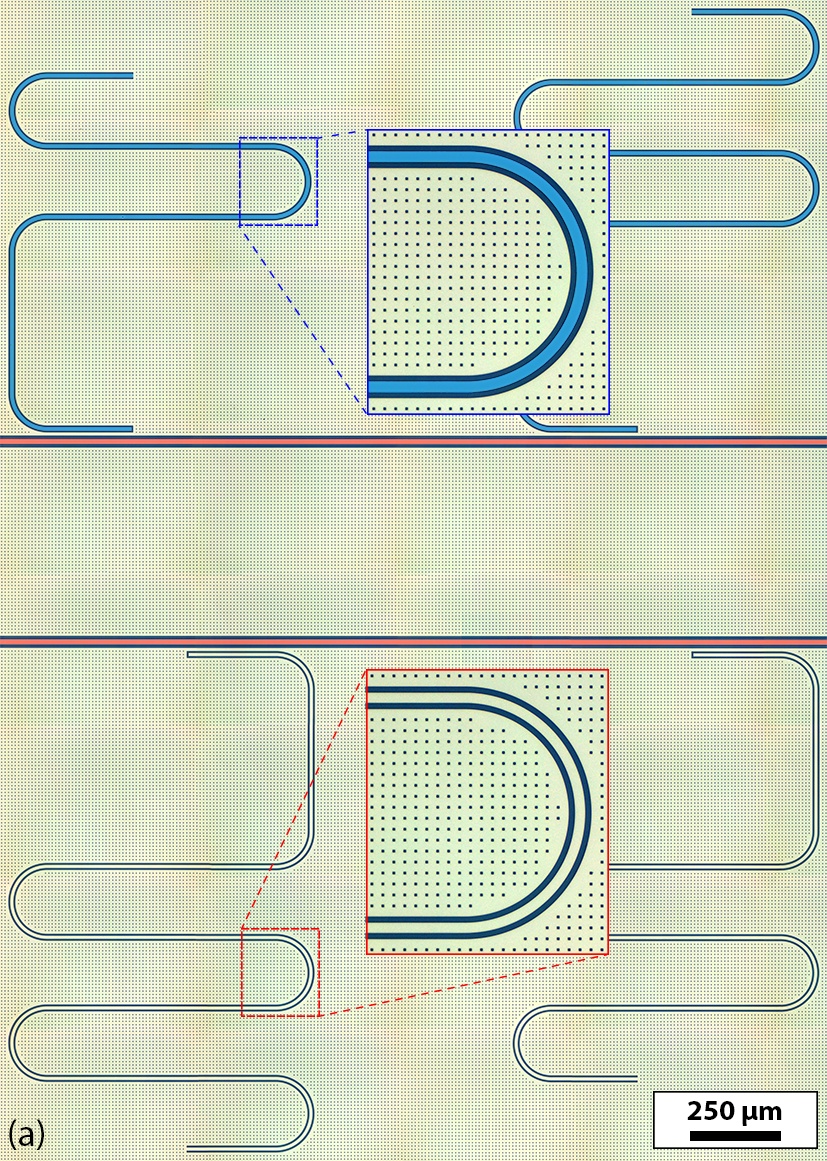}
    \end{minipage}%
    \begin{minipage}{\columnwidth}
        \includegraphics[width=8.5cm]{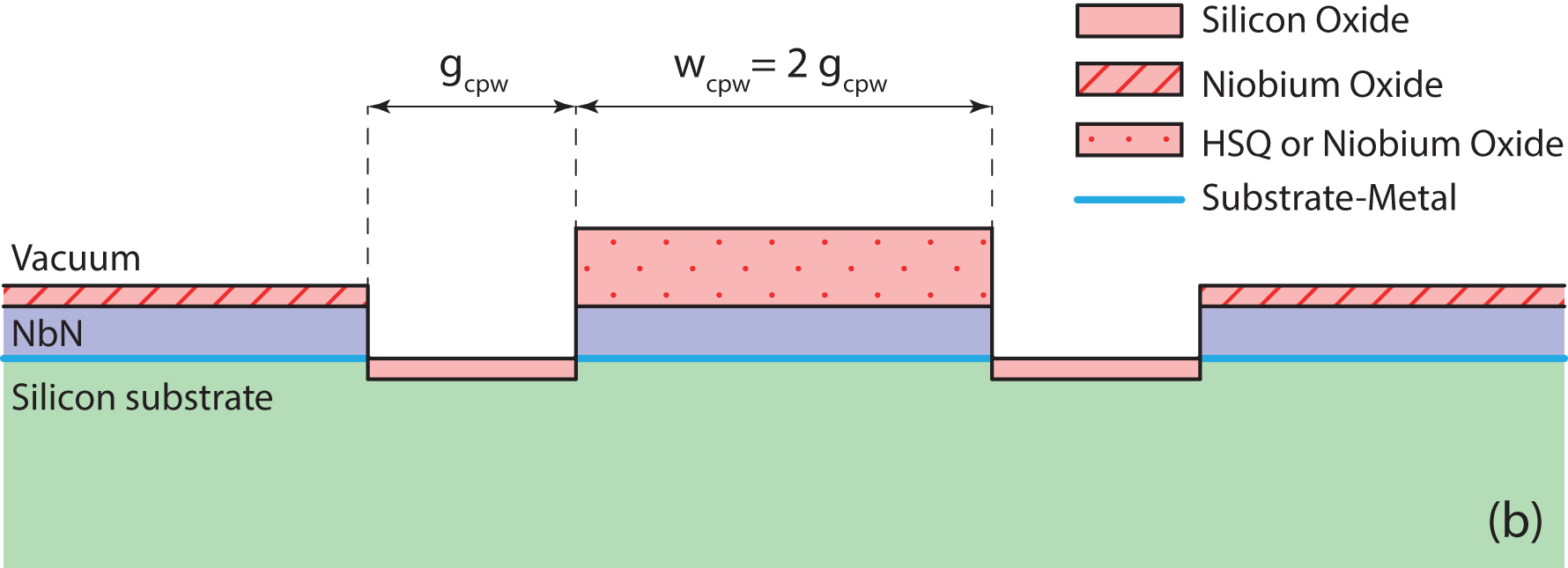}
        
        \vspace{0.5cm}
        
        \includegraphics[width=8.5cm]{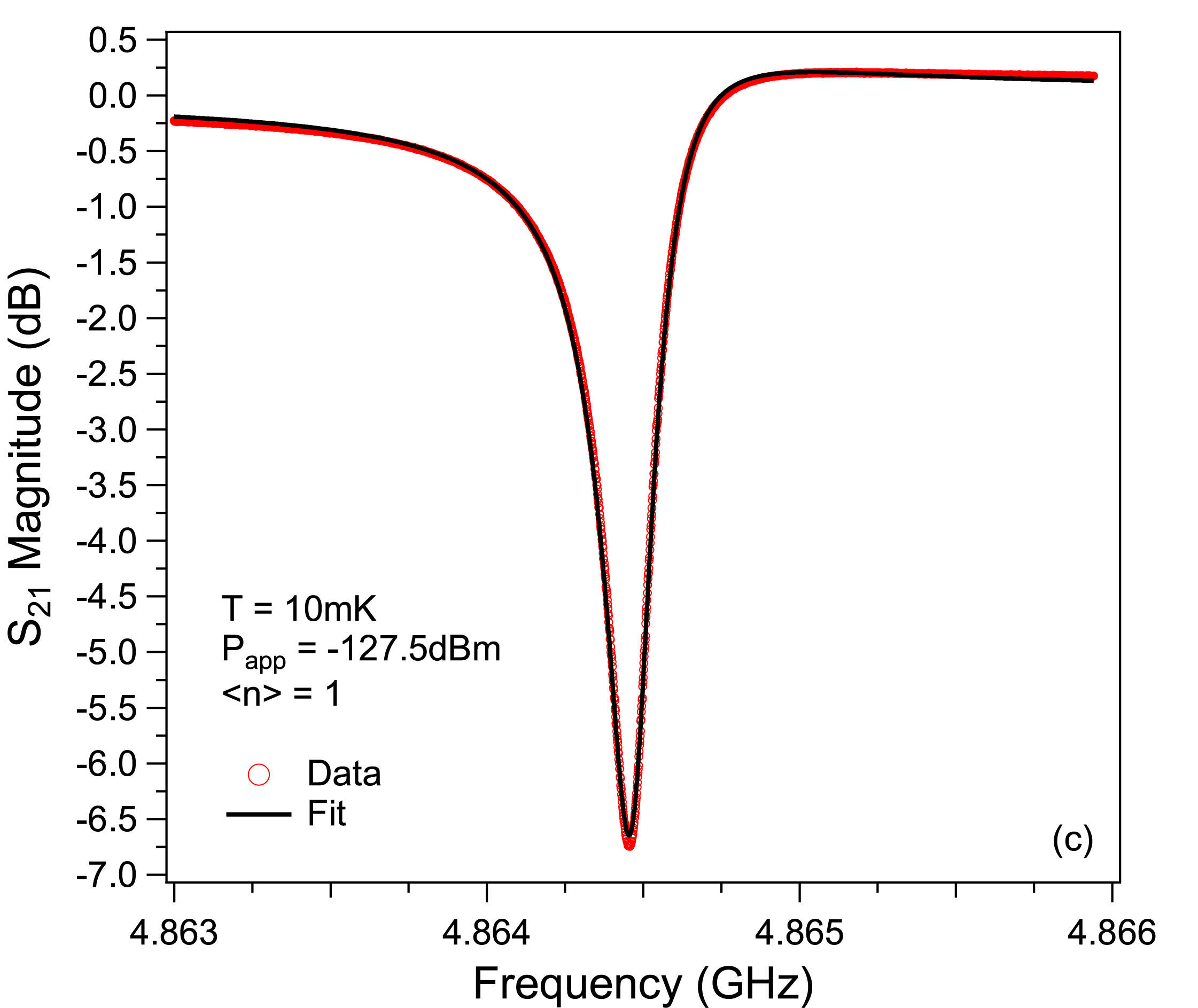}
    \end{minipage}
    
    \caption{\label{fig:device} \textbf{(a)} False-colored optical micrograph of the four resonators used in this work. The resonators are coupled to microwave feed lines (red overlay); the exposed Si substrate, where the NbN has been etched away, is in black. Additionally, HSQ covers the central conductor of the top resonators (cyan overlay). \textbf{(b)} Schematic of the cross-section of the resonators. 
    \newline \textbf{(c)} $S_{21}$ magnitude response of a typical resonator in the single-photon regime (red points). The black line is a fit to determine the resonance parameters.}
\end{figure*}

Several modern circuits rely on superconducting devices with high microwave characteristic impedance and low dissipation. 
High impedance is usually implemented using the kinetic inductance of a chain of Josephson junctions~\cite{Masluk2012Sep,Bell2012Sep,Manucharyan2009Oct} or with sub-micron-width wires made of a disordered superconductor such as NbN~\cite{Niepce2019Apr}, NbTiN~\cite{Samkharadze2016Apr}, or granular Al~\cite{Rotzinger2016Nov,Zhang2019Jan,Grunhaupt2019Apr}. 
Despite being less studied, nanowires have some advantages over junction chains--- high critical current, magnetic-field tolerance~\cite{Samkharadze2016Apr}, strong coupling to zero-point fluctuations of the electric field~\cite{Stockklauser2017Mar,Samkharadze2018Mar}, less stringent constraints on device geometry, and absence of parasitic modes.

Applications of high-impedance devices include qubit architectures such as the fluxonium~\cite{Manucharyan2009Oct,Grunhaupt2019Apr}, which depends on a superinductor (a low-loss inductor with reactive characteristic wave impedance exceeding the resistance quantum, $Z_c>R_Q\sim 6.5\,{\rm k}\Omega$~\cite{Masluk2012Sep,Bell2012Sep,Niepce2019Apr})
and traveling-wave microwave parametric amplifiers~\cite{HoEom2012Jul,Bockstiegel2014Aug,O'Brien2014Oct,White2015Jun,Macklin2015Oct,Vissers2016Jan,Adamyan2016Feb}, relying on the kinetic inductance nonlinearity.
Superconducting disordered nanowires are also interesting for newer types of microwave kinetic-inductance photon detectors (MKIDs)~\cite{Janssen2012May,Schroeder2019Mar} and radio-frequency-readout of superconducting single-photon detectors (SSPDs)~\cite{Schroeder2019Mar,Sinclair2019Aug}.

Dielectric loss and noise associated with two-level systems (TLS) residing in surfaces and interfaces are longstanding problems in superconducting circuits. Specifically, TLS limit the quantum coherence times and lead to parameter fluctuations of superconducting qubits~\cite{Muller2015Jul,Klimov2018Aug,Schlor2019Jan,Burnett2019Jun}. The participation ratios of the losses of the constituent dielectrics can be estimated through electro-magnetic simulation. Traditionally, the air-facing surfaces 
are found to be relatively insignificant, instead, the majority of the loss originates from the substrate--metal and substrate--air interfaces~\cite{Wenner2011Sep,Wang2015Oct,Dial2016Mar,Calusine2018Feb,Woods2019Jul}.
Moreover, for nanowires, the small dimensions exacerbate the TLS contribution to the loss, since the electric field becomes concentrated near the conductor edges. This concentration leads to an increase in the geometric filling factor ($F$) of the lossy dielectric layers compared to that of the loss-less vacuum.
Therefore, it has been demonstrated that TLS remain the dominant loss mechanism even in disordered superconductors with high kinetic inductance, as long as the films are made moderately thin and not excessively disordered~\cite{Niepce2019Apr}.

Across nanowire technologies it becomes necessary to use a spin-on-glass resist to define the sub-micron dimensions. The most prevalent spin-on-glass resist is hydrogen silsesquioxane, HSQ. While HSQ offers unmatched resolution ($\leq\! \SI{10}{\nano\meter}$~\cite{Chen2006Apr}), its structure after development resembles porous amorphous silicon oxide~\cite{Namatsu1998Mar,Liu1998Nov}, which is a well-known host of TLS~\cite{Barends2008Jun}. HSQ is hard to remove after e-beam exposure, and it is therefore often left on top of the finished devices~\cite{Niepce2019Apr}. 

Therefore, when attempting to understand and improve nanowire device performance, we have a rich landscape of small dimensions, disordered superconductors, and spin-on-glass dielectrics, all three of which are quite different from the more commonly used (and consequently well understood) wide ($>\!10\,\mu$m) Al or Nb features fabricated with conventional, removable resists.

In this paper, we explore the geometrical scaling, toward nanowire dimensions, of dielectric losses in microwave resonators. 
We make nominally identical devices with and without spin-on-glass top dielectric and clearly find that in all cases the HSQ makes microwave losses worse. Then, to quantify the loss contributions, we simulate the filling factors and find that due to the ratio of the device dimensions to the London penetration depth, disordered superconductors of small dimensions are not amenable to electrostatic simulations that are traditionally used. 
%
%
To accurately capture the physics, we instead perform 3D finite-element simulations of the current density and electric and magnetic fields at microwave frequencies, from which we extract the various filling factors. This reveals that, while the metal--air interface indeed has a small filling factor, the loss of the HSQ top dielectric is large enough to represent the largest combined loss, in agreement with measurements. 

Combining measurements of the loss and numerical simulation of the filling factors of the different interfaces, we determine the value of the loss tangent of HSQ: $\delta^i_{HSQ} = \SI{8e-3}{}$, i.e.\@ four times that of SiOx~\cite{O'Connell2008Mar,Wang2015Oct,Calusine2018Feb}, which would have been the assumption due to the similarities between spin-on-glass resists and silicon oxide. 

\section{Experimental methods, results}

\begin{figure*}
    \includegraphics[width=8.5cm]{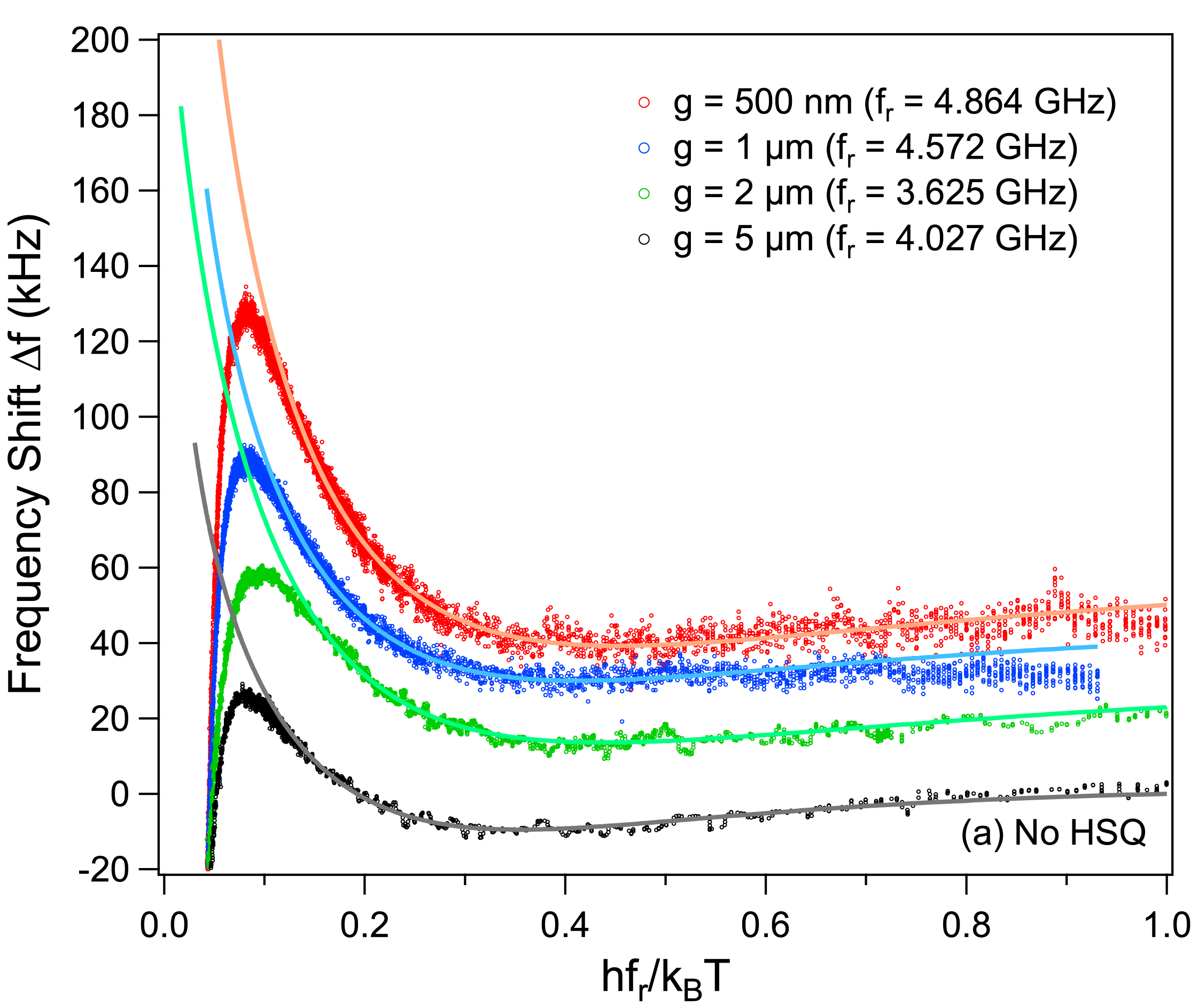}
    \includegraphics[width=8.5cm]{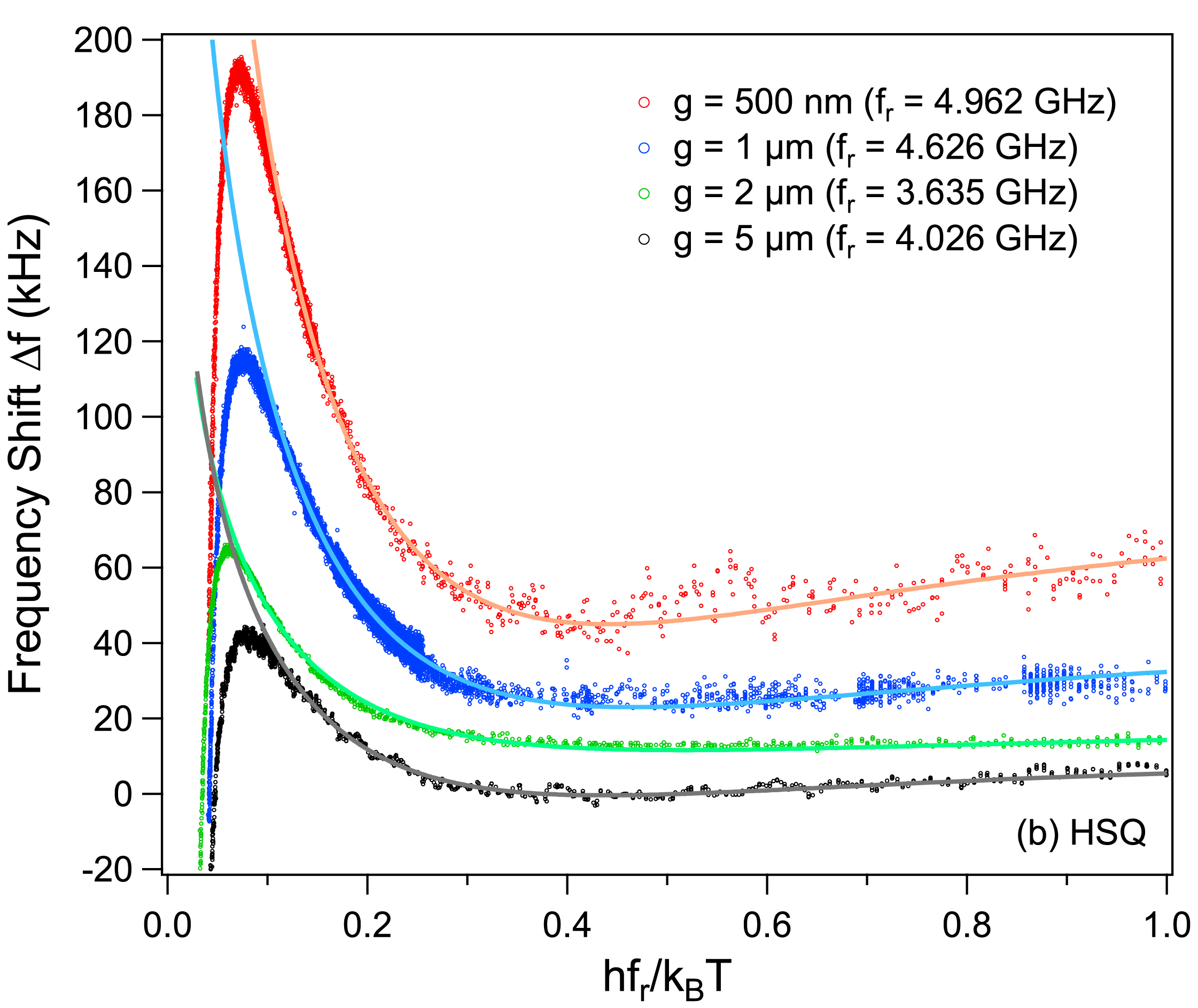}
    
    \caption{\label{fig:Pound} Frequency shift $\Delta f$ as a function of the normalized frequency $f_r$ of the measured resonators without HSQ {\bf (a)} and with HSQ covering the central conductor {\bf (b)}. The data is obtained by applying $P_{app}=-110$~dBm and tracking the changes in resonant frequency against temperature between $\SI{10}{\milli\kelvin}$ and $\SI{1}{\kelvin}$ using the P--FLL.  It is plotted against the natural energy scale of the TLS ($h f_r/k_B T$). The downturn in frequency occurring below $h f_r/k_B T = 0.1$ corresponds to the temperature-dependent kinetic inductance contribution and is not TLS-related.
    For clarity, the curves have been offset by $\SI{15}{\kilo\hertz}$.
    The solid lines are fits to 
    $\Delta f(T) = F_{TLS}\delta^i_{TLS} \left( \ln\left( T/T_0 \right) + \left[ g(T,f_r) - g(T_0,f_r) \right] \right)$~\cite{Gao2008Apr,Lindstrom2009Oct},  
    where $\Delta f (T) = [f_r(T) - f_r(T_0)]/f_r(T_0)$, $T_0$ is a reference temperature, $g(T,f) = \operatorname{Re}\left( \Psi\left( \frac{1}{2} + h f/2\pi i k_B T \right) \right)$, and $\Psi$ is the complex digamma function.
    }
\end{figure*}

In order to study the geometric scaling of dielectric losses, we fabricated NbN coplanar waveguide resonators, with and without HSQ dielectric on top of the center conductor. These devices spanned a range of widths of the center conductor and of the gap between center conductor and ground planes. The gap width ranges from $g_{cpw} = \SI{500}{\nano\meter}$ to $\SI{5}{\micro\meter}$, with the ratio of the gap to the centre conductor kept fixed.
Figure~\ref{fig:device}(a) shows a micrograph of a typical device, and Fig.~\ref{fig:device}(b) shows a sketch of the cross section of the resonators. 

The samples are fabricated on a high-resistivity ($\rho \leq \SI{10}{\kilo\ohm\centi\meter}$) (100) intrinsic silicon substrate. The substrate is dipped for $\SI{30}{\second}$ in a $\SI{2}{\percent}$ hydrofluoric acid (HF) bath to remove the silicon surface oxide. Within $\SI{5}{\minute}$, the wafer is loaded into a UHV sputtering chamber, where a NbN thin-film of thickness $\SI{15}{\nano\meter}$ is deposited by reactive DC magnetron sputtering from a $\SI{99.99}{\percent}$ pure Nb target in a 6:1 Ar:N$_{\text{2}}$ atmosphere at $\SI{12.7}{\micro\bar}$. Next, a $\SI{500}{nm}$-thick layer of PMMA A6 resist is spin-coated and then exposed by electron-beam lithography (EBL) to define the microwave circuitry and resonators. The pattern is developed for $\SI{60}{\second}$ in MIBK:IPA (1:1) and transferred to the film by reactive ion etching (RIE) in a 50:4 Ar:Cl$_{\text{2}}$ plasma at $\SI{50}{\watt}$ and $\SI{10}{\milli\torr}$. In a subsequent EBL exposure, a $\SI{30}{\nano\meter}$ layer of HSQ is first spun and then exposed on the center conductor of half of the microwave resonators such that, after development in a $\SI{2.45}{\percent}$ TMAH solution, each sample has two copies of each design: one covered with HSQ and one without HSQ. 

The samples are wire bonded in a connectorised copper sample box that is mounted onto the mixing chamber of a Bluefors LD250 dilution refrigerator. The inbound microwave signal is attenuated at each temperature stage by a total of $\SI{60}{\decibel}$ before reaching the device under test. Accounting for cable losses and sample-box insertion loss, the total attenuation of the signal reaching the sample is $\SI{70}{\decibel}$. To avoid any parasitic reflections and noise leakage from amplifiers, the transmitted signal is fed through two microwave circulators (Raditek RADI-4.0-8.0-Cryo-4-77K-1WR) and a 4--8 \si{\giga\hertz} band pass filter. Finally, the signal is amplified by a LNF LNC4\_8A HEMT cryogenic amplifier ($\SI{45}{\decibel}$ gain) installed on the 2.8-K stage. Additional amplification is performed at room temperature (Pasternack PE-1522 gain-block amplifiers). 
This measurement environment has been shown to support measurements of resonators with quality factors of several millions~\cite{Burnett2018Mar} and therefore provides an ideal test bench for characterising loss in superconducting microwave resonators.

We study the microwave properties of each of these resonators by measuring the forward transmission ($S_{21}$) response using a Keysight N5249A vector network analyser. When probed with an applied power $P_{app}$, the average energy stored in a resonator of characteristic impedance $Z_c$ and resonant frequency $f_r$ is given by $\left< E_{int} \right> = h f_r \left< n \right> = Z_0 Q_l^2 P_{app} / \pi^2 Z_c Q_c f_r$, where $\left< n \right>$ is the average number of photons in the resonator, $h$ is Planck's constant, $Z_0 = \SI{50}{\ohm}$, and $Q_c$ and $Q_l$ are the coupling and loaded quality factors, respectively. Figure~\ref{fig:device}(c) shows a typical $S_{21}$ magnitude response measured at $\SI{10}{\milli\kelvin}$ and has average photon population $\left< n \right> = 1$. The resonator parameters are extracted by fitting the data with an open-source traceable fit routine~\cite{Probst2015Feb}. 



In order to reliably determine the TLS loss contribution, we measure the resonant frequency of each resonator against temperature between $\SI{10}{\milli\kelvin}$ and $\SI{1}{\kelvin}$~\cite{Gao2008Apr,Lindstrom2009Oct} using a Pound frequency-locked loop (P--FLL). 
The data is shown in Fig.~\ref{fig:Pound}, while the cryogenic microwave setup with the VNA and P--FLL schematics are explained in detail in Ref.~\cite{Niepce2019Apr}. 
This method only probes TLS effects and has the benefit of being sensitive to a wide frequency distribution of TLS. Consequently, the intrinsic loss tangent is robust against spectrally unstable TLS that produce time variations in the quality factor~\cite{Earnest2018Nov}. This allows us to independently determine the intrinsic loss tangent (times the filling factor) $F_{TLS} \delta^i_{TLS}$. The fitted values are presented in Table~\ref{tab:resonators}. 

\begin{table*}
    \centering
    \caption{\label{tab:resonators}Resonator parameters. $F_{TLS}\delta^i_{TLS}$ is obtained from fits of the data in Fig.~\ref{fig:Pound}.}
    
    \begin{ruledtabular}
    \begin{tabular}{cccccc}
        $g_{cpw}$ & $Z_c$ & $f_r$ \footnotesize{(no HSQ)} & $f_r$ \footnotesize{(with HSQ)} & $F_{TLS}\delta^{i}_{TLS}$ \footnotesize{(no HSQ)} & $F_{TLS}\delta^{i}_{TLS}$ \footnotesize{(with HSQ)} \\
         (\si{\micro\meter}) & (\si{\ohm}) & (\si{\mega\hertz}) & (\si{\mega\hertz}) & ($\times 10^{-5}$) & ($\times 10^{-5}$) \\ \colrule
        5 & 207   & 4027 & 4026 & 1.36 & 1.66 \\
        2 & 312   & 3625 & 3635 & 1.60 & 1.87 \\
        1 & 441   & 4572 & 4626 & 1.98 & 2.50 \\
        0.5 & 632 & 4864 & 4962 & 2.74 & 3.92 \\
    \end{tabular}
    \end{ruledtabular}
\end{table*}

\section{Modelling of TLS Loss}

Fig.~\ref{fig:Pound} shows that in our devices, the losses are dominated by TLS, even for thin-film nanowires with widths down to 40~nm. 
In order to accurately account for the individual contributions of all TLS-containing regions of the circuit, we split the dielectric loss into a linear combination of loss tangents each associated with a corresponding filling factor~\cite{Wenner2011Sep,Wang2015Oct,Dial2016Mar,Gambetta2017Jan,Calusine2018Feb},
\begin{equation}
    \label{eq:losstan}
    \frac{1}{Q_{TLS}} = F_{TLS} \delta^i_{TLS} = \sum_{k} F_k \delta^i_k
\end{equation}
where $\delta^i_k$ is the intrinsic loss tangent of region $k$. Additionally, the filling factor of a given TLS host region $k$, of volume $V_k$ and relative permittivity $\varepsilon_k$, is given by %
\begin{equation}
    \label{eq:filling}
    F_k = \frac{U_k}{U_{total}} = \dfrac{\int_{V_k} \varepsilon_{k} \vec{E}^2(\vec{r}) d\vec{r} }{\int_{V} \varepsilon \vec{E}^2(\vec{r}) d\vec{r} }
\end{equation}
where $U_k$ and $U_{total}$ are the electric energy stored in region $k$ and the total electric energy, respectively,  $\vec{E}$ is the electric field, and $\varepsilon$ is the effective permittivity of the entire volume $V$.

Several previous works have studied the loss participation of the different interfaces.
O'Connell et al.~\cite{O'Connell2008Mar} perform low-temperature, low-power microwave measurements, report the intrinsic loss tangent of dielectrics, and interpret their results using a TLS defect model.

Wenner et al.~\cite{Wenner2011Sep} numerically calculate the participation ratios of TLS losses in CPW and microstrip resonators, and find that the losses, at a level of $\delta\sim 5\times 10^{-6}$, predominantly arise due to the substrate--metal (SM) and substrate--air (SA) interfaces, with only a 1-\% contribution from the metal--air (MA) interface. 

Wang et al.~\cite{Wang2015Oct} conduct an experimental and numerical study of losses in Al transmon qubits and attribute the dominant loss to surface dielectrics, consistent with the TLS loss model. In a literature study of transmons made with the standard lift-off process, they find a seemingly universal value $\tan \delta \sim 2.6 \times 10^{-3}$. We note that the spread between data points pertaining to different devices is within the range of temporal variation, due to spectrally unstable TLS, recently reported in both qubit $T_1$~\cite{Burnett2019Jun} and resonator $Q$~\cite{Earnest2018Nov}.

Dial et al.~\cite{Dial2016Mar} experimentally study 3D transmon qubits, with results consistent with the SM and SA interfaces being the dominant contributors to loss. 

Calusine et al.~\cite{Calusine2018Feb,Woods2019Jul} trench the substrate of TiN resonators, achieving a mean low-power quality factor of $~3\times 10^6$, and demonstrate agreement with a finite-element electrostatic simulation of dielectric loss.

\section{Filling Factor Simulations}

\begin{figure}
    \includegraphics[width=8.5cm]{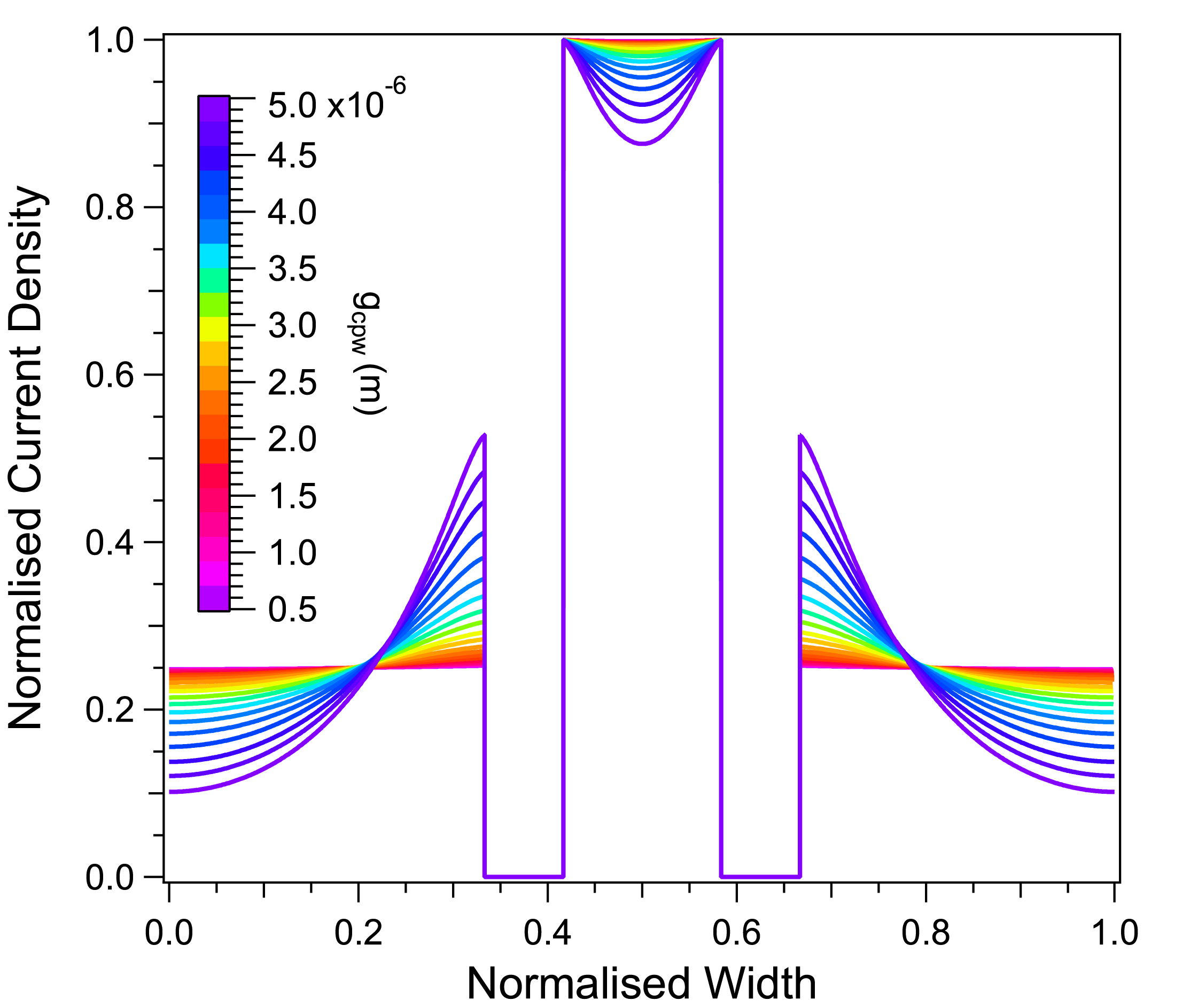}
    
    \caption{\label{fig:current} Simulated normalised current density inside the superconductors, extracted along a line half-way inside it (half the thickness), for all simulated values $g_{cpw}$ in the $\SI{500}{\nano\meter}$ to $\SI{5}{\micro\meter}$ range. }
\end{figure}

\begin{figure}
    \includegraphics[width=8.5cm]{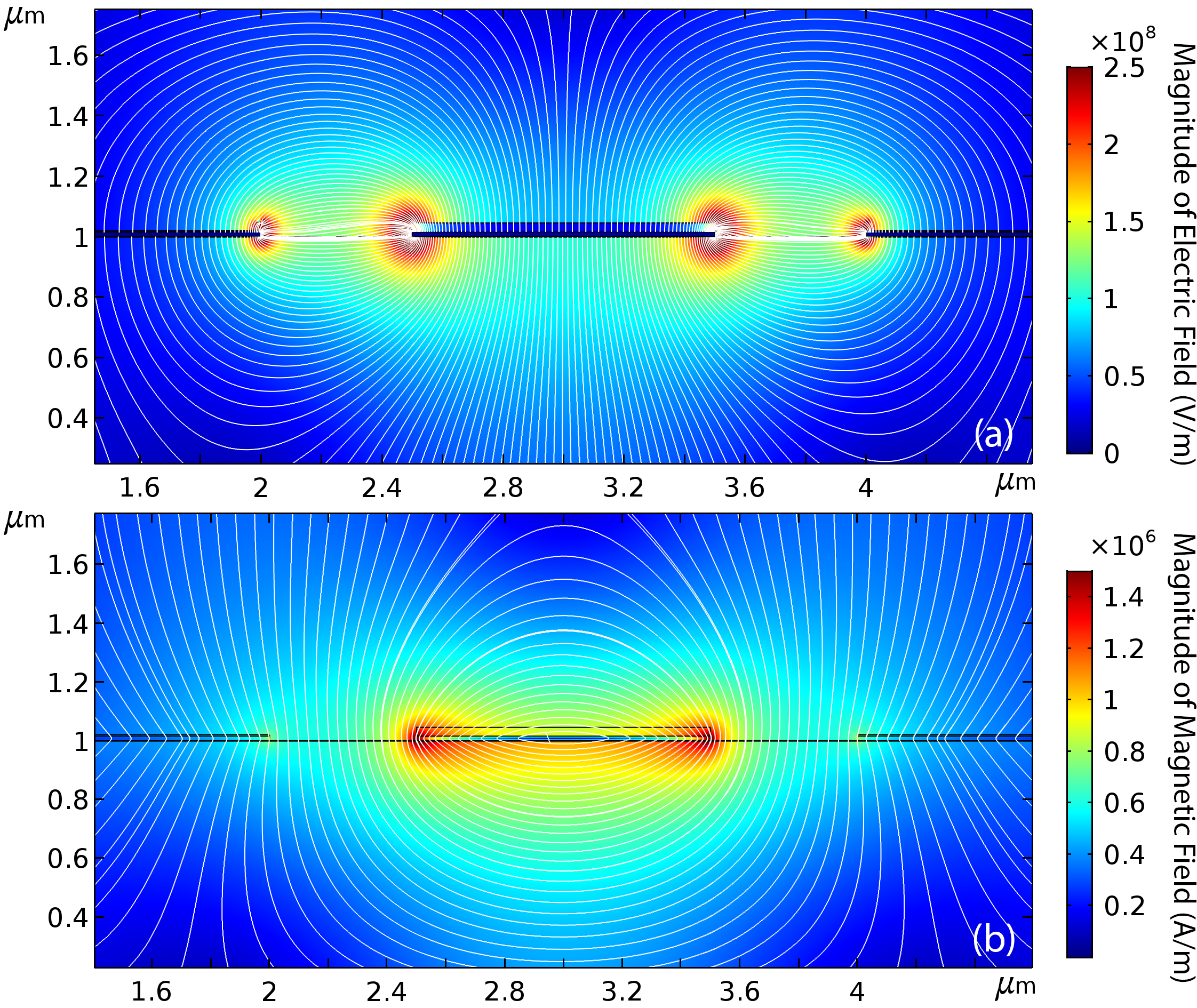}
    
    \caption{\label{fig:comsol} Magnitude and field lines of the simulated electric \textbf{(a)} and magnetic fields \textbf{(b)} for a cross section of the resonators with HSQ covering the central conductor. The permittivity in Eq.~(\ref{eq:london_sim}), with $\omega/2\pi = \SI{5}{\giga\hertz}$, is given as an input to the Comsol Multiphysics simulation tool.}
\end{figure}

\begin{figure*}
    \includegraphics[width=8.5cm]{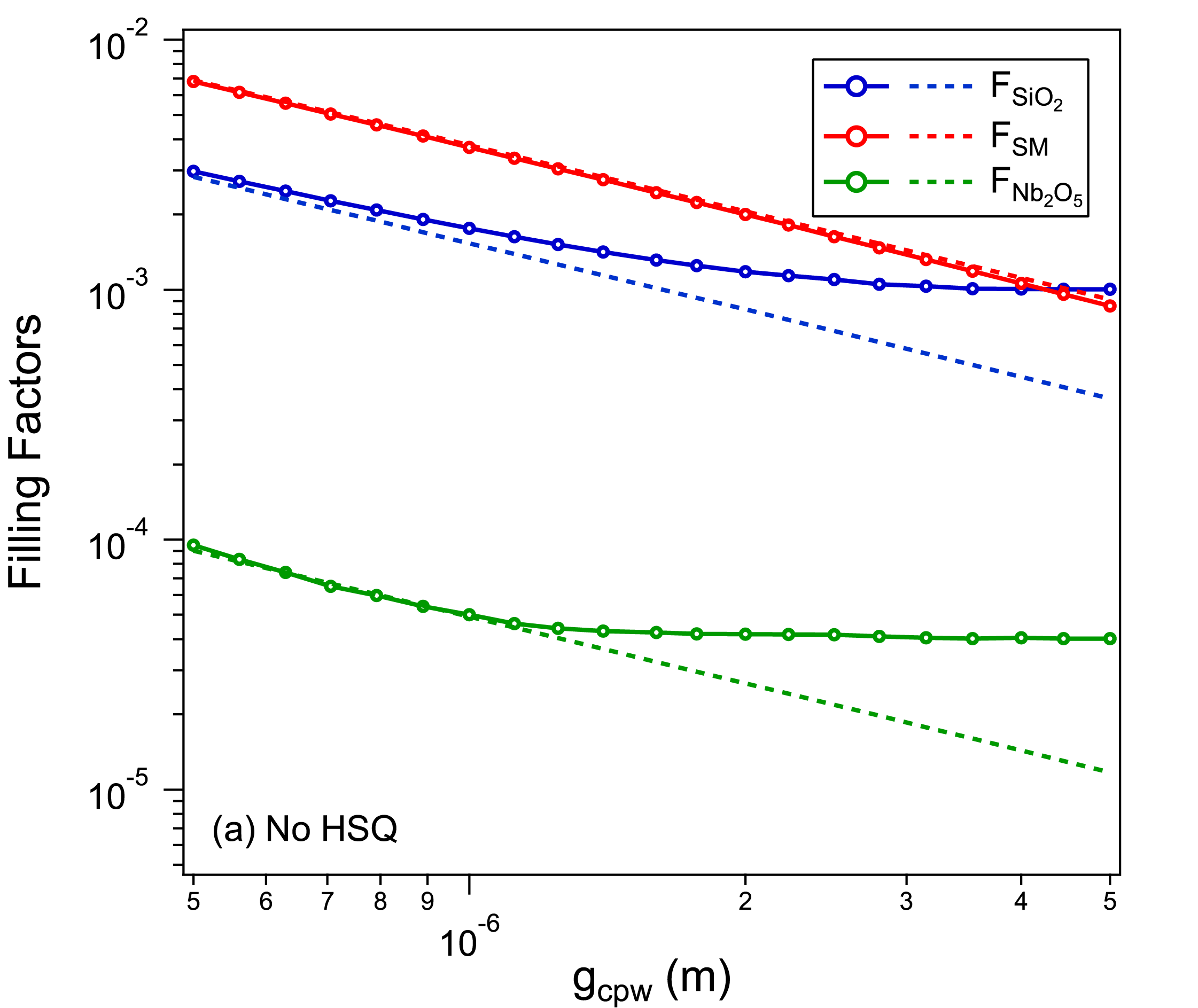}
    \includegraphics[width=8.5cm]{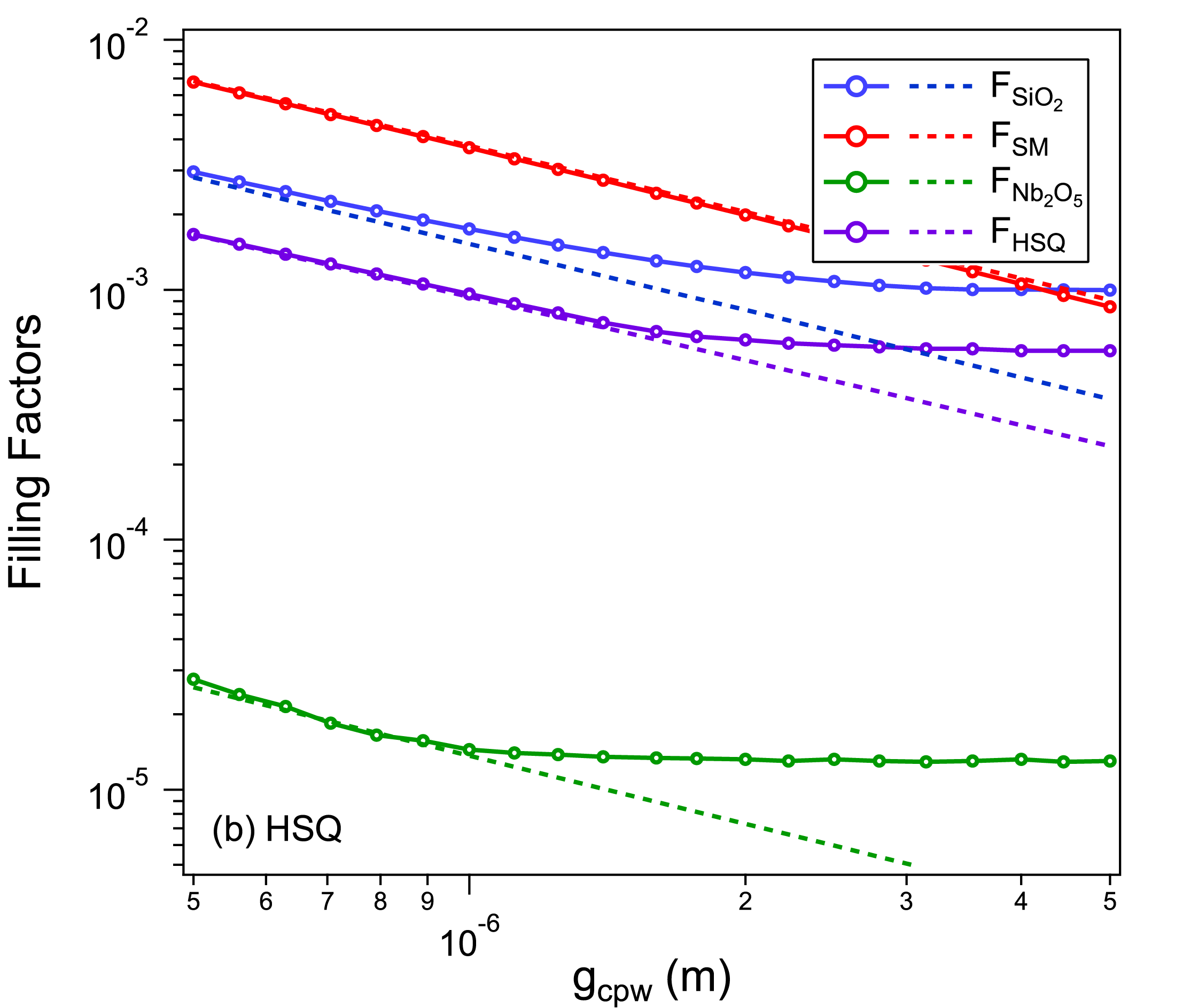}
    
    \caption{\label{fig:filling} Simulated filling factors $F$ as a function of the co-planar waveguide gap $g_{cpw}$ for resonators without HSQ \textbf{(a)} and with HSQ covering the central conductor \textbf{(b)}. The dashed lines represent the incorrect $F$ obtained with electrostatic simulations.}
\end{figure*}



In order to analyse dielectric and interfacial losses in our devices, and in particular to identify those from the HSQ top dielectric, we perform electro-magnetic simulations (with and without the HSQ layer) in Comsol Multiphysics for a wide range of resonator geometries. A sketch of the cross-section of the simulated structures is shown in Fig.~\ref{fig:device}(b). 
The simulation parameters for the constituent materials are as follows: the SA interface is modelled as a $\SI{5}{\nano\meter}$ thick layer of \chem{SiO_2}~\cite{Morita1990Aug} with relative permittivity $\varepsilon_r(\chem{SiO_2}) = 4.2$. The MA interface consists of a $\SI{5}{\nano\meter}$ thick layer of \chem{Nb_2 O_5}~\cite{Henry2017Jun} with relative permittivity $\varepsilon_r(\chem{Nb_2 O_5}) = 33$~\cite{Kaiser2010Jun,Romanenko2017Dec}. The SM interface is modelled by a $\SI{2}{\nano\meter}$ thick layer inside the substrate ($\varepsilon_r(\chem{SM}) = \varepsilon_r(\chem{Si})= 11.7$)~\cite{Calusine2018Feb}. Finally, the HSQ region has a thickness of $\SI{30}{\nano\meter}$ and relative permittivity $\varepsilon_r(\chem{HSQ}) = 3$~\cite{Liu1998Nov}. Because \chem{Nb_2 O_5} requires several days to achieve any meaningful thickness~\cite{Henry2017Jun}, it is assumed that no \chem{Nb_2 O_5} is present underneath the HSQ. Therefore, on the samples without HSQ, \chem{Nb_2 O_5} resides on both the central conductor and ground planes, whereas on the samples with HSQ, \chem{Nb_2 O_5} is present only on the ground planes.

The superconductor part of the structure requires extra care to simulate accurately: strongly disordered superconductors, like NbN, have an extremely small electron mean free path $l$ (on the order of $\SI{0.5}{\nano\meter}$ and smaller~\cite{Chockalingam2008Jun}) and are therefore in the local dirty limit~\cite{Dressel2013Jul}. In this limit, several quantities become dependent on the mean free path and need to be adjusted from their BCS values~\cite{Gor'kov1959Dec,tinkham}. Most importantly for this study, the magnetic penetration depth in disordered superconductors and at zero temperature becomes
\begin{equation}
    \label{eq:dirty_lambda0}
    \lambda_{dirty}(0) = \lambda_L(0) \sqrt{\dfrac{\xi_0}{l}} = \sqrt{ \dfrac{\hbar}{\pi\mu_0 \Delta_0 \sigma_n} }
\end{equation}
\noindent where $\lambda_L(0)$ is the London penetration depth at $T=\SI{0}{\kelvin}$, $\xi_0$ is the BCS coherence length, $\hbar$ is the reduced Planck constant, $\mu_0$ is the vacuum permeability, $\Delta_0$ is the superconducting gap at zero temperature, and $\sigma_n$ is the normal-state conductivity. Additionally, the temperature dependence of the penetration depth is given by
\begin{equation}
    \label{th:eq:dirty_lambda_T}
   \dfrac{\lambda_{dirty}(T)}{\lambda_{dirty}(0)} = \left[ \dfrac{\Delta(T)}{\Delta_0}  \tanh \left( \dfrac{\Delta(T)}{2 k_B T} \right) \right]^{-1/3}
\end{equation}

By measuring the resistance vs.\@ temperature of our NbN thin films, we find $T_c = \SI{7.20}{\kelvin}$ and $\sigma_n = \SI{1.32e5}{\siemens\per\meter}$ (measured at the onset of the superconducting transition). Using $\Delta_0 = 2.08 k_B T_c$~\cite{Mondal2011Jan}, we obtain $\lambda_{dirty} = \SI{987}{\nano\meter} \simeq \SI{1}{\micro\meter}$, which is comparable to the lateral dimension of our resonators. 

Consequently, it is not sufficient to approximate the current density in our NbN devices as a surface density, since magnetic fields significantly penetrate the superconductor. This is in contrast to resonators made of a conventional superconductor such as aluminium ($\lambda_L(0) \simeq \SI{30}{\nano\meter}$~\cite{Maloney1972May}) or niobium ($\SI{100}{\nano\meter}$~\cite{Langley1991Jul}). In a similar way, it is insufficient to assume a uniform current distribution in the superconductor when the resonator dimensions are smaller than $\lambda_L(T)$.

Therefore, a static solution of Maxwell's equations is insufficient here, in particular for the wider geometries. 
Instead we need to solve the Maxwell--London equations, at the relevant frequency of the alternating current, in order to accurately simulate the densities of the current and electromagnetic fields. We achieve this in a 3D finite-element simulator by considering the superconductor as an environment with a complex permittivity~\cite{Vendik1998May,Javadzadeh2013Apr},

\begin{equation}
    \label{eq:london_sim}
    \varepsilon_r(\omega,T) = \varepsilon_0 - \dfrac{1}{\omega^2 \mu_0 \lambda_{dirty}(T)^2} - j\dfrac{\sigma_1(\omega,T)}{\omega}
\end{equation}
\noindent where $\sigma_1(\omega,T)$ is the real part of the Mattis--Bardeen conductivity.

The meshing of the simulated structure has to be carefully optimised due the vast difference of length scales within the resonator structure (widths, thicknesses, and also the wavelength). The simulation mesh is manually defined using Comsol's swept mesh functionality and consists of rectangular elements. Rectangular elements are preferred over the more standard tetrahedral elements to avoid poor meshing quality inherent to high-aspect ratio tetrahedrons. The edge length of each element is varied from $\SI{3}{\nano\meter}$ to $\SI{100}{\nano\meter}$, with smaller elements close to the regions of interest (superconducting thin-film and dielectric layers). Due to memory constraints, however, the edge length alongside the wave propagation direction is kept constant to $\SI{100}{\nano\meter}$ and only a short section of co-planar waveguide is simulated ($l_{cpw} = \SI{4}{\micro\meter}$). A relative tolerance of $\SI{1e-5}{}$ was found as a good compromise between the accuracy of the converged solution and the duration of the simulation.

Figures~\ref{fig:current}--\ref{fig:comsol} show the simulated current density and electric and magnetic fields, respectively, for a cross section of a resonator with $g_{cpw} = \SI{500}{\nano\meter}$. 
From the electric fields, we calculate the filling factor of each region using Eq.~(\ref{eq:filling}) and present the result in Fig.~\ref{fig:filling}. Additionally, Fig.~\ref{fig:filling} shows filling factors calculated by means of electrostatic simulation to highlight the significant deviation from the Maxwell--London simulation results for $w_{cpw} > \lambda_L$. 

\begin{figure}
    \includegraphics[width=8.5cm]{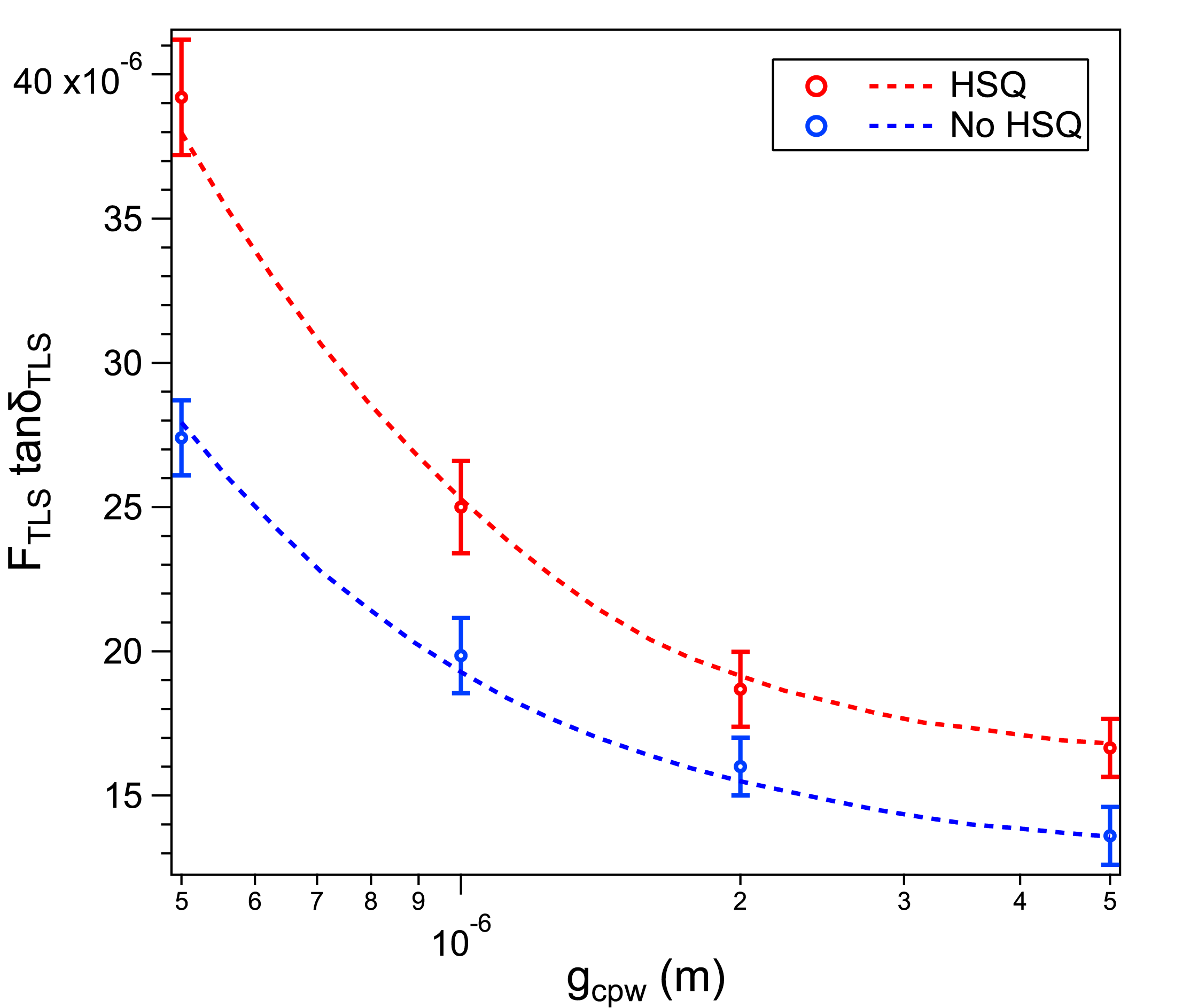}

    \caption{\label{fig:Ftand_sim} Total TLS loss $F_{TLS}\delta^i_{TLS}$ vs.\@ gap width $g_{cpw}$ of the co-planar waveguide for all four measured resonators. The $F_{TLS}\delta^i_{TLS}$ values are determined from fits of the $\Delta f(T)$ data in Fig.~\ref{fig:Pound}
    ---see Table~\ref{tab:resonators}.
    The error bars represent two standard deviations of uncertainty (95\% confidence interval). The dashed lines are fits to Eq.~(\ref{eq:losstan}) using the simulated filling factors $F_{TLS}$ shown in Fig.~\ref{fig:filling}.}
\end{figure}

Using these simulated filling factors, we can fit  Eq.~(\ref{eq:losstan}) to the experimental results in Table~\ref{tab:resonators}---see Fig.~\ref{fig:Ftand_sim}---and in this way determine the intrinsic loss tangent of each lossy region. These results are summarised in Table~\ref{tab:tand}. 

\begin{figure*}
    \includegraphics[width=8.5cm]{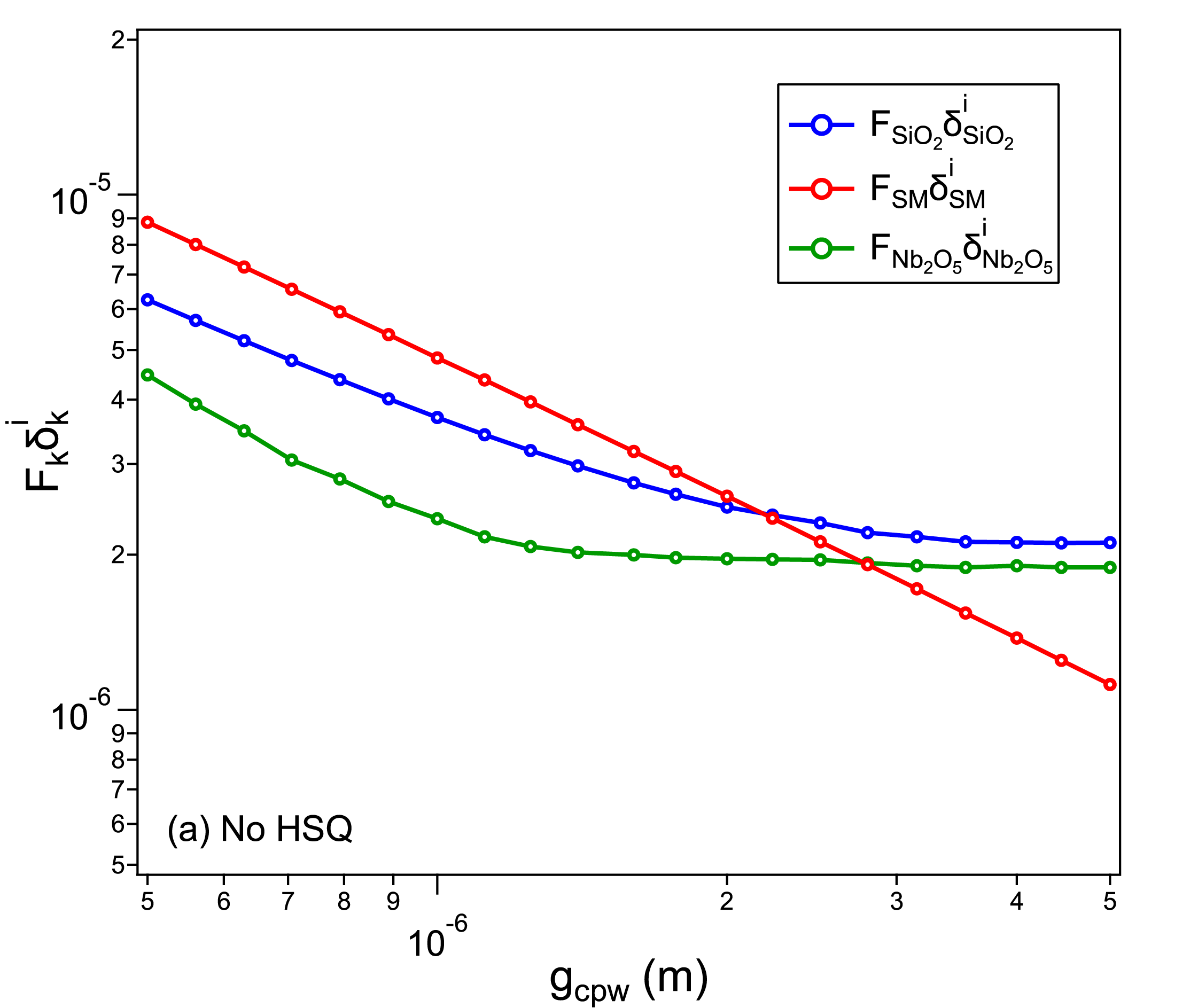}
    \includegraphics[width=8.5cm]{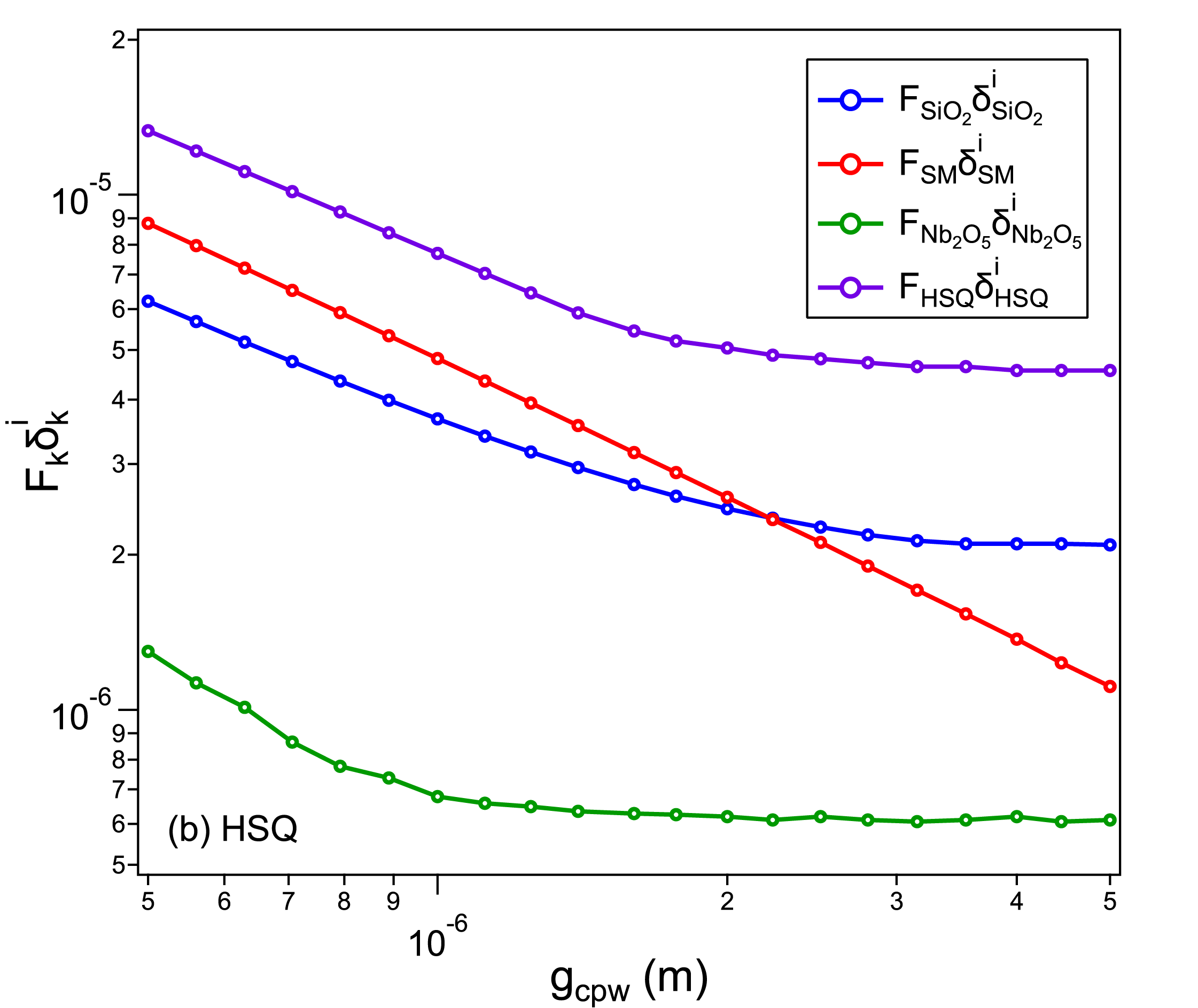}
    
    \caption{\label{fig:ftand} Contribution of each individual lossy region for resonators without HSQ \textbf{(a)} and with HSQ \textbf{(b)} covering the central conductor. }
\end{figure*}

\vfill 

\begin{table}[h!]
    \centering
    \caption{\label{tab:tand} Fitted loss tangents of the different lossy regions. The values are obtained from fits to Eq.~(\ref{eq:losstan}) using the simulated filling factors.}
    
    \begin{ruledtabular}
    \begin{tabular}{ccc}
        Region                      & Symbol                        & Value\\ \colrule
        HSQ                         & $\delta^i_{HSQ}$              & $\SI{8.0e-3}{}$ \\
        Substrate-Metal Interface   & $\delta^i_{SM}$               & $\SI{1.3e-3}{}$ \\
        Niobium Oxide               & $\delta^i_{\chem{Nb_2O_5}}$   & $\SI{4.7e-2}{}$ \\
        Silicon Oxide               & $\delta^i_{\chem{SiO_2}}$     & $\SI{2.1e-3}{}$ \\
    \end{tabular}
    \end{ruledtabular}
\end{table}

\section{Discussion}



Our results are consistent with values found by other groups in similar types of devices~\cite{O'Connell2008Mar,Kaiser2010Jun,Wang2015Oct,Calusine2018Feb}. However, we emphasise that the fabrication of our devices was not focused on minimising the influence of TLS. 

We find the intrinsic loss tangent for HSQ to be $\delta^i_{HSQ} = \SI{8.0e-3}{}$. Paired with the relatively large filling factor of the HSQ region, this makes HSQ the dominant contribution to the loss for all dimensions, as highlighted in Fig.~\ref{fig:ftand}; and for a given dimension, $F_{TLS} \delta^i_{TLS}$ is systematically higher for the sample covered with HSQ, as shown in Fig.~\ref{fig:Ftand_sim}.
These results confirms that the porous amorphous silicon oxide structure of developed HSQ~\cite{Namatsu1998Mar,Liu1998Nov} is a major source of dielectric loss, and therefore, a process that allows for the removal of the HSQ mask would lead to significant improvements in device performance.

\section{Conclusion}

In conclusion, we  fabricated and measured co-planar waveguide resonators with dimensions ranging from $g_{cpw} = \SI{5}{\micro\meter}$ down to $\SI{500}{\nano\meter}$ in order to study the geometric dependence of TLS loss. Using 3D finite-element electro-magnetic simulations we calculated the relative contributions of the different sources of TLS loss. Such simulations provide a valuable tool to predict the performance of superconducting resonators and other superconducting quantum devices. 

Additionally, by comparing resonators with the central conductor covered by HSQ and resonators without HSQ, we were able to extract the intrinsic loss tangent of this dielectric: $\delta^i_{HSQ} = \SI{8.0e-3}{}$. 

\begin{acknowledgments}

We acknowledge support in the device fabrication from the Chalmers Nanofabrication Laboratory staff. The authors are grateful to Philippe Tassin for granting access to his nodes in the Chalmers C3SE computational cluster where our simulations were performed. This research has been supported by funding from the Swedish Research Council and Chalmers Area of Advance Nanotechnology. In addition, J.J.B acknowledges financial support from the Industrial Strategy Challenge Fund Metrology Fellowship as part of the UK government’s Department for Business, Energy and Industrial Strategy.
\end{acknowledgments}

\bibliography{refs.bib}

\end{document}